# Calculating Kolmogorov Complexity from the Transcriptome Data

Panpaki Seekaki[1] and Norichika Ogata[1]

[1] Nihon BioData Corporation, 3-2-1 Sakado, Takatsu-ku, Kawasaki, Kanagawa, Japan
`norichik@nbiodata.com`

**Abstract.** Information entropy is used to summarize transcriptome data, but ignoring zero count data contained them. Ignoring zero count data causes loss of information and sometimes it was difficult to distinguish between multiple transcriptomes. Here, we estimate Kolmogorov complexity of transcriptome treating zero count data and distinguish similar transcriptome data.

## 1    Introduction

When environmental conditions change abruptly, living cells must coordinate adjustments in their genome expression to accommodate the changing environment [1]. It is possible that the degree of change in the environment affects the degree of change in the gene expression pattern. However, it is difficult to completely understand the amount of change that occurs in the transcriptome, given that this would involve thousands of gene expression measurements. A previous study defined transcriptome diversity as the Shannon entropy of its frequency distribution, and made it possible to express the transcriptome as a single value. Dimensionality Reduction methods e.g. Principle component analysis and t-SNE had been used to transcriptome analyses [2, 3], but biological meanings of value of results of these methods was unclear. Transcriptome diversity also reduces the dimensions of transcriptome data and the biological meaning of transcriptome diversity is clear. The first research on transcriptome diversity was performed with human tissues [1], and later research compared cancer cells with normal cells [4]. In plant research, transcriptome diversity has been used to compare several wounded leaves [5]. Transcriptome diversity also used for measurement of cellular dedifferentiation [6]. Our Previous study compared transcriptome diversity between cells cultured in vitro in media supplemented with several concentrations of phenobarbital, to investigate how the amount of environmental change (in terms of drug concentration) affects the amount of transcriptome change and multi-stability of the genome expression system was indicated by an observation of hysteretic phenomenon of transcriptome diversities [7].

Information entropy of transcriptome were described in a previous study [1], the transcriptomes of each tissue as a set of relative frequencies, $P_{ij}$, for the $i$th gene ($i =$ 1, 2, …, $g$) in the $j$th tissue ($j$ = 1, 2, …, $t$); and then quantified transcriptome diversity using an adaptation of Shannon's entropy formula:

$$H_{ij} = -\sum_{i=1}^{g} P_{ij} \log_2(P_{ij})$$

Transcriptome data obtained using RNA-seq contains zero count data. In Shannon entropy, these zero count data were ignored. There is possibility that this founder mental character of information entropy makes it difficult to distinguish similar transcriptomes, ignoring zero count data causes loss of information. We hypothesized that developing a transcriptome-summarizing method treating zero count makes distinguish similar transcriptomes possible.

The elementary theories of Shannon information and Kolmogorov complexity have a common purpose [8]. Kolmogorov complexity is the minimum number of bits from which a particular message or file can effectively be reconstructed. Since zero count data contained transcriptome data have a message, Kolmogorov complexity treat zero count data.

Here, we developed a Kolmogorov complexity calculating method from transcriptome data containing zero count data. Similar transcriptomes that were not distinguished using Information entropy were distinguished using Kolmogorov complexity. Monte Carlo simulation indicated that Kolmogorov complexity does not detect genes order in transcriptome data files.

## 2    Materials and Methods

### 2.1    Transcriptome data

Transcriptome sequence data from two cellular culture conditions were obtained DDBJ SRA. Short-read data have been deposited in the DNA Data Bank of Japan (DDBJ)'s Short Read Archive, under project ID DRA002853. We choose DRX025343 and DRX025346 as similar transcriptomes that were not well distinguished using Information entropy. These two data sets contain three replicates, respectively. Short-read sequences were mapped to an annotated silkworm transcript sequence as previously described [7]. I show the first four lines of the file as an example (Table 1).

**Table 1.** An example of transcriptome data.

| rownames(data) | DRX3_1 | DRX3_2 | DRX3_3 | DRX6_1 | DRX6_2 | DRX6_3 |
|---|---|---|---|---|---|---|
| BGIBMGA000001 | 6 | 9 | 7 | 6 | 6 | 5 |
| BGIBMGA000002 | 39 | 18 | 25 | 22 | 36 | 23 |
| BGIBMGA000003 | 1 | 0 | 0 | 0 | 0 | 0 |

DRX3_1 is a replicate of DRX025343.

## 2.2 Calculation of Kolmogorov Complexity

Calculation of Kolmogorov Complexity was performed using UNIX and R 3.0.2 [9]. At first in R, we converted reads count data to RPM data and converted RPM data to rounded RPM data. Reads count data was named "for_R_count.txt".

```
data<-read.table("for_R_count.txt",header=T,row.names=1)
out_f<-"for_R_RPM.txt"
param1 <- 1000000
norm_factor <- param1/colSums(data)
out <- sweep(data, 2, norm_factor, "*")
tmp <- cbind(rownames(data), out)
write.table(tmp,out_f,sep="\t",append=F,          quote=F,
row.names=F, col.names=T)
data<-read.table("for_R_RPM.txt",header=T,row.names=1)
out_f <- "temp1"
write.table(round(data), out_f, sep="\t", append=F, quote=F,
row.names=T, col.names=F)
```

Then we made rounded RPM data in UNIX.

```
head -n 1 for_R_RPM.txt > temp2
cat temp2 temp1 > for_R_RPMround.txt
```

We converted rounded RPM data from decimal number to 22-digit binary number For example, "33" was converted "0000000000000000100001".

```
more temp2 | tr "\t" "\n" | awk '{print "more temp1 | awk
@{print $"NR"}@ > "$1}' | grep -v rowname | tr "@" "'" >
do_single.sh
sh do_single.sh
more temp2 | tr "\t" "\n" | grep -v rowname | awk '{print
"more  "$1"  |  awk  @{print  \"echo  \\\"obase=2;ibase=10;
\"$1\"\\\"  |  bc  |  awk  V{printf(\\\"%022d\\\n\\\",$1)}V >>
2shin_"$1".text\"}@ | tr \"V\" \"@\" >> do_conv_10_2.sh"}' |
tr "@" "'" > make_do_conv_10_2.sh
sh make_do_conv_10_2.sh
sh do_conv_10_2.sh
```

We compressed rounded RPM data (22-digit binary number) using zip command.

```
more temp2 | tr "\t" "\n" | grep -v rownames | awk '{print
"zip 2shin_"$1".zip 2shin_"$1".text"}' > do_zip.sh
sh do_zip.sh
```

We compared file sizes between rounded RPM data (22-digit binary number) and zip compressed rounded RPM data (22-digit binary number).

```
ls -al | grep 2shin | ls -al | grep 2shin | grep zip | awk
'{print $9"\t"$5}' > temp3
ls -al | grep 2shin | ls -al | grep 2shin | grep text | awk
'{print $9"\t"$5}' > temp4
paste temp3 temp4 | tr "\." "\t" | sed -e 's/2shin_//g' |
awk '{print $1"\t"($3/$6*1.00)"\t"$3"\t"$6}' > temp5
echo "sampleID@kc@zip_filesize@raw_filesize" | tr "@" "\t" >
temp6
cat                  temp6                  temp5                  >
for_R_sampleID_kc_ZIPfilesize_RAWfilesize.txt
```

These codes were written in a file named "cal_KC.sh". I show the first four lines of the output file as an example (Table 2).

**Table 2.** An example of the Output file.

| sampleID | kc | zip_filesize | raw_filesize |
|----------|-----------|--------------|--------------|
| DRX3_1 | 0.0692269 | 23283 | 336329 |
| DRX3_2 | 0.0692209 | 23281 | 336329 |
| DRX3_3 | 0.0697502 | 23459 | 336329 |

## 2.3 Monte Carlo Simulation

We choose DRX025343 and DRX025346 for MC simulation since these transcriptomes were similar and were not clearly distinguished using information entropy. We extracted data of these two samples in UNIX.

```
more  for_R_RPMround.txt  | cut  -f  1,5,6,7,20,21,22  >
for_R_RPMround_ovs+.txt
more for_R_RPMround_ovs+.txt | grep -v row | cut -f 2- >
temp01.txt
```

We obtained random numbers for each gene in R. We performed MC simulation 13 times.

```
a<-runif(14623,0,1)
b<-runif(14623,0,1)
c<-runif(14623,0,1)
d<-runif(14623,0,1)
e<-runif(14623,0,1)
f<-runif(14623,0,1)
```

```
g<-runif(14623,0,1)
j<-runif(14623,0,1)
h<-runif(14623,0,1)
i<-runif(14623,0,1)
k<-runif(14623,0,1)
l<-runif(14623,0,1)
m<-runif(14623,0,1)
n<-runif(14623,0,1)
x<- data.frame(a,b,c,d,e,f,g,h,i,j,k,l,m,n)
write.table(x,"runif.txt")
```

We gave random numbers to each gene and sorted genes in the order of given random numbers.

```
more runif.txt | tr -d "\"" | tr " " "\t" | cut -f 2- | tail
-n 14623 > unif_table.txt
   paste temp01.txt unif_table.txt | awk '{print $7"\t"$0}' |
sort | cut -f 2,3,4,5,6,7 > sorted_7.txt
   paste temp01.txt unif_table.txt | awk '{print $8"\t"$0}' |
sort | cut -f 2,3,4,5,6,7 > sorted_8.txt
   paste temp01.txt unif_table.txt | awk '{print $9"\t"$0}' |
sort | cut -f 2,3,4,5,6,7 > sorted_9.txt
   paste temp01.txt unif_table.txt | awk '{print $10"\t"$0}' |
sort | cut -f 2,3,4,5,6,7 > sorted_10.txt
   paste temp01.txt unif_table.txt | awk '{print $11"\t"$0}' |
sort | cut -f 2,3,4,5,6,7 > sorted_11.txt
   paste temp01.txt unif_table.txt | awk '{print $12"\t"$0}' |
sort | cut -f 2,3,4,5,6,7 > sorted_12.txt
   paste temp01.txt unif_table.txt | awk '{print $13"\t"$0}' |
sort | cut -f 2,3,4,5,6,7 > sorted_13.txt
   paste temp01.txt unif_table.txt | awk '{print $14"\t"$0}' |
sort | cut -f 2,3,4,5,6,7 > sorted_14.txt
   paste temp01.txt unif_table.txt | awk '{print $15"\t"$0}' |
sort | cut -f 2,3,4,5,6,7 > sorted_15.txt
   paste temp01.txt unif_table.txt | awk '{print $16"\t"$0}' |
sort | cut -f 2,3,4,5,6,7 > sorted_16.txt
   paste temp01.txt unif_table.txt | awk '{print $17"\t"$0}' |
sort | cut -f 2,3,4,5,6,7 > sorted_17.txt
   paste temp01.txt unif_table.txt | awk '{print $18"\t"$0}' |
sort | cut -f 2,3,4,5,6,7 > sorted_18.txt
   paste temp01.txt unif_table.txt | awk '{print $19"\t"$0}' |
sort | cut -f 2,3,4,5,6,7 > sorted_19.txt
```

To calculate Kolmogorov complexity of 13 simulations, we modified "cal_KC.sh" and named "cal_KC_mod.sh". The following "cal_KC_mod.sh" is displayed.

```
    more temp2 | tr "\t" "\n" | awk '{print "more temp1 | awk
@{print $"NR"}@ > "$1}' | grep -v rowname | tr "@" "'" >
do_single.sh
    sh do_single.sh
    more temp2 | tr "\t" "\n" | grep -v rowname | awk '{print
"more  "$1"  |  awk  @{print  \"echo  \\\"obase=2;ibase=10;
\"$1"\\\"" | bc | awk V{printf(\\\"%022d\\\\n\\\",$1)}V >>
2shin_"$1".text\"}@ | tr \"V\" \"@\" >> do_conv_10_2.sh}' |
tr "@" "'" > make_do_conv_10_2.sh
    sh make_do_conv_10_2.sh
    sh do_conv_10_2.sh
    more temp2 | tr "\t" "\n" | grep -v rownames | awk '{print
"zip 2shin_"$1".zip 2shin_"$1".text"}' > do_zip.sh
    sh do_zip.sh
    ls -al | grep 2shin | ls -al | grep 2shin | grep zip | awk
'{print $9"\t"$5}' > temp3
    ls -al | grep 2shin | ls -al | grep 2shin | grep text | awk
'{print $9"\t"$5}' > temp4
    paste temp3 temp4 | tr "\." "\t" | sed -e 's/2shin_//g' |
awk '{print $1"\t"($3/$6*1.00)"\t"$3"\t"$6}' > temp5
    echo "sampleID@kc@zip_filesize@raw_filesize" | tr "@" "\t" >
temp6
    cat                temp6                temp5                >
for_R_sampleID_kc_ZIPfilesize_RAWfilesize.txt
    rm 2shin*
    head -n 1 for_R_RPM.txt | tr "\t" "\n" | grep -v rownames |
awk '{print "rm "$1}' > clean_single_lanes.sh
    sh clean_single_lanes.sh
    rm clean_single_lanes.sh
    rm temp*
    rm do*
    rm make*
    rm *.text
    echo "look for_R_sampleID_kc_ZIPfilesize_RAWfilesize.txt"
    echo "end \(^q^)"
```

We calculated Kolmogorov complexity using "cal_KC_mod.sh".

```
    mkdir unif_07
    cd unif_07/
    paste ../rownames.txt ../sorted_7.txt > temp_sorted_7.txt
    cat ../header.txt temp_sorted_7.txt > for_R_RPMround.txt
    more for_R_RPMround.txt | tail -n 14623 > temp1
    head -n 1 for_R_RPMround.txt > temp2
```

```
cp ../cal_KC_mod.sh ./
sh cal_KC_mod.sh
```

We named the codes below "temp02.txt" following.

```
mkdir unif_08
cd unif_08/
paste ../rownames.txt ../sorted_8.txt > temp_sorted_8.txt
cat ../header.txt temp_sorted_8.txt > for_R_RPMround.txt
more for_R_RPMround.txt | tail -n 14623 > temp1
head -n 1 for_R_RPMround.txt > temp2
cp ../cal_KC_mod.sh ./
sh cal_KC_mod.sh
```

We calculated the rested 11 simulations.

```
 more temp02.txt | tr "\n" "@" | sed -e 's/08/09/g' | sed -e
's/8/9/g' | tr "@" "\n" > do_cal_KC_mod.sh
 more temp02.txt | tr "\n" "@" | sed -e 's/08/10/g' | sed -e
's/8/10/g' | tr "@" "\n" >> do_cal_KC_mod.sh
 more temp02.txt | tr "\n" "@" | sed -e 's/08/11/g' | sed -e
's/8/11/g' | tr "@" "\n" >> do_cal_KC_mod.sh
 more temp02.txt | tr "\n" "@" | sed -e 's/08/X/g' | sed -e
's/8/X/g' |  sed -e 's/X/12/g' |  tr "@"  "\n" >>
do_cal_KC_mod.sh
 more temp02.txt | tr "\n" "@" | sed -e 's/08/X/g' | sed -e
's/8/X/g' |  sed -e 's/X/13/g' |  tr "@"  "\n" >>
do_cal_KC_mod.sh
 more temp02.txt | tr "\n" "@" | sed -e 's/08/X/g' | sed -e
's/8/X/g' |  sed -e 's/X/14/g' |  tr "@"  "\n" >>
do_cal_KC_mod.sh
 more temp02.txt | tr "\n" "@" | sed -e 's/08/X/g' | sed -e
's/8/X/g' |  sed -e 's/X/15/g' |  tr "@"  "\n" >>
do_cal_KC_mod.sh
 more temp02.txt | tr "\n" "@" | sed -e 's/08/X/g' | sed -e
's/8/X/g' |  sed -e 's/X/16/g' |  tr "@"  "\n" >>
do_cal_KC_mod.sh
 more temp02.txt | tr "\n" "@" | sed -e 's/08/X/g' | sed -e
's/8/X/g' |  sed -e 's/X/17/g' |  tr "@"  "\n" >>
do_cal_KC_mod.sh
 more temp02.txt | tr "\n" "@" | sed -e 's/08/X/g' | sed -e
's/8/X/g' |  sed -e 's/X/18/g' |  tr "@"  "\n" >>
do_cal_KC_mod.sh
```

```
  more temp02.txt | tr "\n" "@" | sed -e 's/08/X/g' | sed -e
's/8/X/g'  |  sed  -e  's/X/19/g'  |   tr  "@"  "\n"  >>
do_cal_KC_mod.sh
  sh do_cal_KC_mod.sh
```

## 2.4 Additional example using Mammalian Cells

Validate relationships information entropy and Kolmogorov complexity; we cultured adherent mammalian cells in several concentrations of drug and calculated information entropy and Kolmogorov complexity of transcriptomes. Experiments were performed as previously described [7]. We compared the magnitude of the correlation between drug concentrations and information entropy and the correlation between drug concentrations and Kolmogorov complexity in R.

```
  data<-
read.table("temp04_condtion_sampleID_kc_ie.txt",header=T)
  attach(data)
  model<-lm(drug_c~kc+ie)
  summary(model)
```

## 3 Results and Discussion

### 3.1 Comparison of information entropy and Kolmogorov complexity

We calculated Kolmogorov complexity from the transcriptome data deposited in the DNA Data Bank of Japan (DDBJ)'s Short Read Archive, under project ID DRA002853. Previous study presented a comparison between the amount of environmental change and the amount of transcriptome change. In that research, the amount of environmental change was defined by drug concentration and the amount of transcriptome change was defined by information entropy (transcriptome diversity) of transcriptomes. Here, we change definition of the amount of transcriptome change, to Kolmogorov complexity from information entropy. We compared the relationship between Kolmogorov complexity of transcriptome and drug concentration and the relationship between information entropy of transcriptome and drug concentration (see Table 3, Fig. 1).

**Table 3.** Drug concentration and information entropy and Kolmogorov complexity.

| DDBJ_ID | pb | hyst | ie | kc |
|---|---|---|---|---|
| DRR027746 | 0 | F | 9.958201 | 0.067109883 |
| DRR027747 | 0 | F | 10.39145 | 0.068986023 |
| DRR027748 | 0 | F | 10.55041 | 0.069116847 |
| DRR027752 | 0.25 | F | 10.17702 | 0.068641122 |
| DRR027753 | 0.25 | F | 10.3602 | 0.068572737 |

| DRR027754 | 0.25 | F | 10.43436 | 0.069113874 |
| DRR027758 | 0 | T | 10.12837 | 0.067062311 |
| DRR027759 | 0 | T | 9.751655 | 0.065798667 |
| DRR027760 | 0 | T | 10.26499 | 0.067508303 |
| DRR027761 | 0.25 | T | 9.349972 | 0.063761971 |
| DRR027762 | 0.25 | T | 10.22082 | 0.067332879 |
| DRR027763 | 0.25 | T | 9.110797 | 0.063574655 |
| DRR027755 | 1 | F | 8.171864 | 0.060137544 |
| DRR027756 | 1 | F | 7.569893 | 0.059911575 |
| DRR027757 | 1 | F | 8.002486 | 0.055995766 |
| DRR027764 | 2.5 | F | 7.681749 | 0.058249512 |
| DRR027765 | 2.5 | F | 7.334465 | 0.057211837 |
| DRR027766 | 2.5 | F | 8.059749 | 0.055249473 |
| DRR027767 | 12.5 | F | 8.004586 | 0.058974992 |
| DRR027768 | 12.5 | F | 7.519157 | 0.056935322 |
| DRR027769 | 12.5 | F | 7.931476 | 0.058570626 |

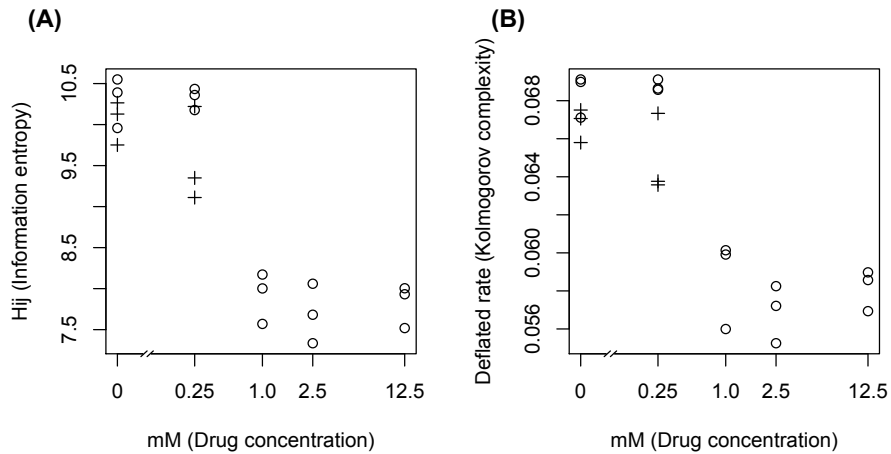

**Fig. 1. Scatter plots of the amount of environmental change vs the amount of transcriptome change.** Transcriptome data deposited in the DNA Data Bank of Japan (DDBJ)'s Short Read Archive, under project ID DRA002853 were used. (A) Scatter plot of drug concentration vs information entropy of transcriptome. (B) Scatter plot of drug concentration vs Kolmogorov complexity of transcriptome.

Information entropy and Kolmogorov complexity of cells cultured in media containing 0.25 mM phenobarbital after previous cultivation (cultivation for 80 hours in MGM-450 insect medium without phenobarbital, followed by cultivation for 10 hours in 1.0 mM phenobarbital-supplemented MGM-450 insect medium) (see Fig. 1 plot "+") were different from cells cultured in media containing 0.25 mM phenobarbital after previous cultivation (cultivation for 80 hours in MGM-450 insect medium without phenobarbital) (see Fig. 1 plot "○"). The hysteretic phenomenon of transcriptome was reproduced using Kolmogorov complexity. In plot (A), using information entropy, a range of plot "+" and a range plot "o" overlapped. In plot (B), using Kolmogorov complexity, a range of plot "+" and a range plot "o" did not overlap. These result indicated that the similar transcriptomes that were not well distinguished using Information entropy were well distinguished using Kolmogorov complexity.

### 3.2    Kolmogorov complexity and the order of genes of transcriptomes

Generally, genes expression data are saved as count tables (see Table 1). The order of those tables is changed easily since that is determined by the short-read sequences mapping process and the annotation process. Generally, genes expression data are saved as count tables (See Table 1). The order of those tables is changed easily since that is determined by the short-read sequences mapping process and the annotation process. The order of genes also has any message as well as zero count data and have a possibility to affect Kolmogorov complexity. Monte Carlo simulations were performed to know the amount of effect of the order of genes to Kolmogorov complexity. We simulated 13 time, we gave random numbers for each gene and randomized the order of genes of transcriptomes. As a result, the order of genes in transcriptome does not correlate Kolmogorov complexity (see Table 4, Fig. 2).

**Table 4.** Kolmogorov complexity of 13 MC simulations.

| MC | DRX3_1 | DRX3_2 | DRX3_3 | DRX6_1 | DRX6_2 | DRX6_3 |
|----|--------|--------|--------|--------|--------|--------|
| 1 | 0.0692269 | 0.0692209 | 0.0697502 | 0.0643685 | 0.0681773 | 0.0639493 |
| 2 | 0.0691614 | 0.0693428 | 0.0694617 | 0.0643983 | 0.0681654 | 0.0641782 |
| 3 | 0.0692952 | 0.0692477 | 0.0696729 | 0.0645529 | 0.0681862 | 0.0641842 |
| 4 | 0.0694439 | 0.0692952 | 0.0696283 | 0.0642317 | 0.068097 | 0.063872 |
| 5 | 0.0691496 | 0.0693815 | 0.0697026 | 0.0642704 | 0.0680792 | 0.0643864 |
| 6 | 0.0694707 | 0.0694588 | 0.0695093 | 0.0643864 | 0.0679335 | 0.0641426 |
| 7 | 0.0695063 | 0.0693577 | 0.0695034 | 0.0645261 | 0.0679989 | 0.064095 |
| 8 | 0.0694112 | 0.0691942 | 0.0695837 | 0.0640861 | 0.0681505 | 0.0640504 |
| 9 | 0.0692209 | 0.0694201 | 0.0696699 | 0.0641842 | 0.0679602 | 0.0641634 |
| 10 | 0.069432 | 0.0695093 | 0.0697383 | 0.0643477 | 0.0679573 | 0.0638868 |
| 11 | 0.0693666 | 0.0693042 | 0.0699256 | 0.0645023 | 0.0678978 | 0.0636579 |
| 12 | 0.0694796 | 0.069545 | 0.069322 | 0.0643328 | 0.0681713 | 0.0642763 |
| 13 | 0.0694766 | 0.0693636 | 0.069542 | 0.0643091 | 0.0680286 | 0.0641574 |

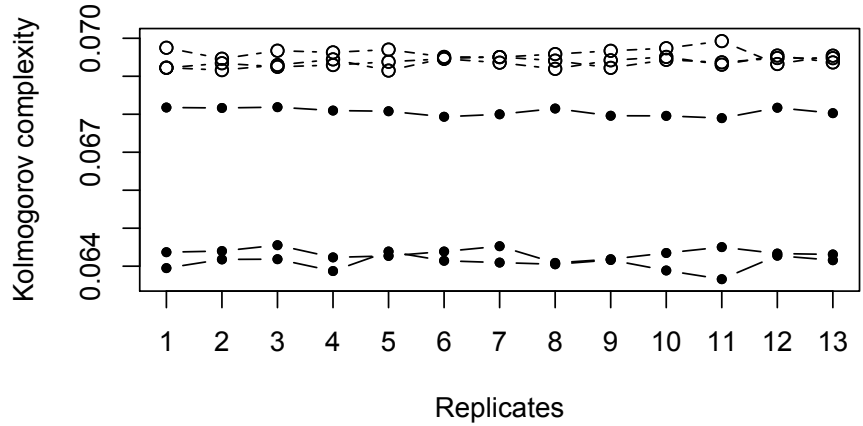

**Fig. 2. Kolmogorov complexity of genes order randomized transcriptome.** Gene orders of 6 transcriptomes were randomized by Monte Carlo simulation.

### 3.3    Additional example using Mammalian Cells

To validate relationships information entropy and Kolmogorov complexity, we cultured adherent mammalian cells in several concentrations of drug and calculated information entropy and Kolmogorov complexity of transcriptomes (Table 5). We named table 5, "temp04_condtion_sampleID_kc_ie.txt".

**Table 5.** Relative drug concentration and information entropy and Kolmogorov complexity.

| drug_c | ie | kc |
|--------|----------|-----------|
| 0 | 11.79063 | 0.0607232 |
| 0 | 11.71702 | 0.0609433 |
| 0 | 11.8423 | 0.0610302 |
| 0.1 | 11.92365 | 0.0610032 |
| 0.1 | 11.82417 | 0.0609453 |
| 0.1 | 11.82852 | 0.0610109 |
| 1 | 11.51067 | 0.0597326 |
| 1 | 11.84796 | 0.0612445 |
| 1 | 11.98672 | 0.0611866 |
| 10 | 11.79985 | 0.0610824 |
| 10 | 11.79642 | 0.0611094 |
| 10 | 11.84944 | 0.0613025 |

We compared the correlation between information entropies and drug concentrations and the correlation between Kolmogorov complexities and drug concentrations using Fitting Linear Models in R.

```
Call:
lm(formula = drug_c ~ kc + ie)

Residuals:
     Min       1Q   Median       3Q      Max
-0.57247 -0.27188 -0.03609  0.25045  0.54018

Coefficients:
            Estimate Std. Error t value Pr(>|t|)
(Intercept)  -28.083     18.967  -1.481   0.1728
kc          1080.540    558.570   1.934   0.0851 .
ie            -3.175      1.980  -1.603   0.1434
---
Signif. codes: 0.001 '**' 0.01 '*' 0.05 '.' 0.1 '' 1

Residual standard error: 0.4065 on 9 degrees of freedom
Multiple R-squared:  0.2939,  Adjusted R-squared:  0.137
F-statistic: 1.873 on 2 and 9 DF, p-value: 0.2089
```

The correlation between Kolmogorov complexities and drug concentrations was stronger than that of information entropies (see Fig.3).

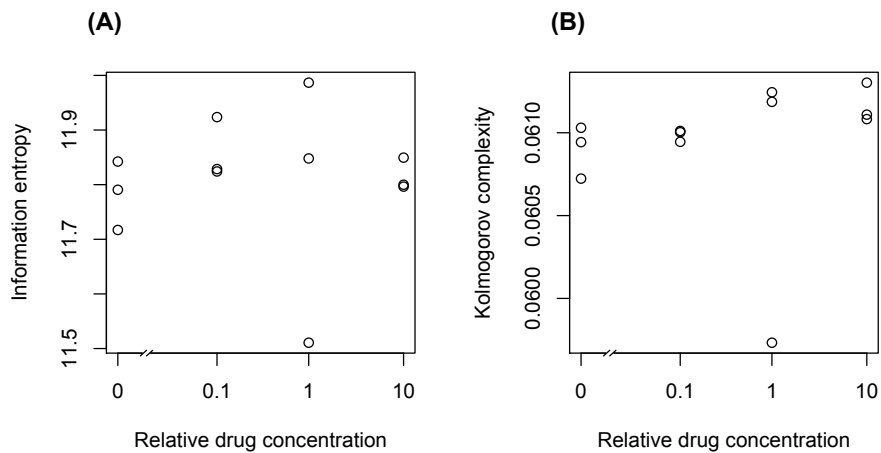

**Fig. 3. Scatter plots of the amount of environmental change vs the amount of transcriptome change in mammalian cells.** (A) Scatter plot of relative drug concentration vs information entropy of transcriptome. (B) Scatter plot of relative drug concentration vs Kolmogorov complexity of transcriptome.

# 4    Conclusions

Transcriptome measurement technologies and applications of information theories to transcriptome data allow studying the cellular systems. Especially, Shannon's theory of information explained several phenomena [10]. In this study, to expand the use of information theories, we improved weakness "Ignoring Zero Counts" of Shannon's information entropy in transcriptome analyses using Kolmogorov complexity. Kolmogorov complexity treat zero count data in transcriptome data and distinguished similar transcriptomes that were not well distinguished using information entropy.